# Hard X-ray helical dichroism of disordered molecular media

Jérémy R. Rouxel*,1,2, Benedikt Rösner*,3, Dmitry Karpov3, Camila Bacellar1, Giulia F. Mancini4, Francesco Zinna5,6, Dominik Kinschel1, Oliviero Cannelli1, Malte Oppermann1, Cris Svetina3, Ana Diaz3, Jérôme Lacour5, Christian David3, and Majed Chergui*,1

1 Ecole Polytechnique Fédérale de Lausanne, Laboratoire de Spectroscopie Ultrarapide (LSU) and Lausanne Centre for Ultrafast Science (LACUS),CH-1015 Lausanne, Switzerland
2 Univ Lyon, UJM-Saint-Etienne, CNRS, Graduate School Optics Institute, Laboratoire Hubert Curien UMR 5516, Saint-Etienne F-42023, France
3 Paul Scherrer Institut, 5232 Villigen PSI, Switzerland
4 Physics Department - University of Pavia, Via Agostino Bassi 6, 27100 Pavia (PV), Italy
5 Department of Organic Chemistry, University of Geneva, quai Ernest Ansermet 30, 1211Geneva 4, Switzerland
6 Dipartimento di Chimica e Chimica Indistriale, Università Di Pisa, via Moruzzi 13, Pisa,56124 Italy

E-mail : jeremy.rouxel@univ-st-etienne.fr; benedikt.roesner@psi.ch; majed.chergui@epfl.ch

**Chirality is a structural property of molecules lacking mirror symmetry[1] that has strong implications in diverse fields, ranging from life to materials sciences. Established spectroscopic methods that are sensitive to chirality, such as circular dichroism (CD), exhibit weak signal contributions on an achiral background. Helical dichroism (HD), which is based on the orbital angular momentum (OAM) of light, offers a new approach to probe molecular chirality,[2,3] but it has never been demonstrated on disordered samples. Furthermore, in the optical domain the challenge lies in the need to transfer the OAM of the photon to an electron that is localized on an Å-size orbital. Here, we overcome this challenge using hard X-rays with spiral Fresnel zone, which can induce an OAM. We present the first HD spectra of a disordered powder sample of enantiopure molecular complexes of $[Fe(4,4'-diMebpy)_3]^{2+}$ at the iron K-edge (7.1 keV) with OAM-carrying beams. The HD spectra exhibit the expected inversions of signs switching from a left to a right helical wave front or from an enantiomer to the other. The asymmetry ratios for the HD spectra are within one to five percent for OAM beams with topological charges of one and three. These results open a new window into the studies of molecular chirality and its interaction with the orbital angular momentum of light.**

Chiral molecules exist in two mirror-image non-superimposable conformations that are called enantiomers. Although chemically identical, the two enantiomers can have dramatically different chemical reactions and biological functions. Therefore, the asymmetric synthesis and the detection of a single type of enantiomer of chiral compounds are topics of crucial importance, due to their numerous applications, ranging from life to materials sciences.

The two most commonly used methods to distinguish enantiomers are Optical Rotatory Dispersion (ORD) and Circular Dichroism (CD). ORD is the rotation of the plane of linear polarization of visible-ultraviolet light induced by the propagation through a chiral medium, while CD denotes the difference in absorption of left- and right-handed circularly polarized light in chiral molecular systems. As CD vanishes in achiral media, it is a unique probe of molecular chirality. Both ORD and CD are based on interactions of the investigated chiral molecules with the spin angular momentum (SAM) of light. The SAM of light can take values of +1 or -1, corresponding to the left and right circular polarization states respectively. They rely on magnetic dipole interactions, a higher order term in the multipolar expansion of the interaction Hamiltonian of light with matter. As a consequence, CD signals correspond to typically ~ 0.1% of the total linear absorption.

Besides a SAM, photons can possess an unbound OAM, which corresponds to the spatial phase distribution of the wave front.[2,4] In contrast to CD, transferring the OAM within a photon beam to an electron in an atom or molecule is extremely challenging and hard to detect.[3] The main reason for this lies in the need to transfer the phase of an electromagnetic wave that can easily extend over several hundreds of nanometers, to an electron that is typically localized in an Ångström-size orbital. Additionally, the centre of the investigated molecule or atom is never perfectly aligned with the center of the phase vortex and it has been shown that the relative position of the material system with respect to the OAM center is of crucial importance for the transitions amplitudes.[5] Despite these difficulties, progress has been made in the observation of momentum transfer from an OAM beam to bound electrons,[6] and recently to photoelectrons that were ionized from Rydberg states by an OAM-carrying infrared beam.[7] A scheme using OAM beams to trap and discriminate optically chiral molecules has also been proposed.[8]

HD is the differential absorption between two linearly-polarized OAM beams with left and right helicity. Similar to the case of the electronic wavefunction, a non-zero OAM field is obtained when the expectation value of the OAM operator $\mathbf{L} = -i\hbar\, \mathbf{r}\times\mathbf{p}$ is non-zero. In particular, fields that are eigenvectors of this operator with a non-zero eigenvalue carry an OAM (or topological charge) of an integer value L. These fields are typically described as the Laguerre-Gauss or the hypergeometric-Gaussian modes of the electromagnetic radiation (see Supplementary Materials §S1) and display a phase twist along the axis of propagation of $e^{iL\varphi}$ where $\varphi$ is the angular coordinate in cylindrical coordinates. The description of the SAM and the OAM of vortex beams have long been discussed using either classical[9] or quantum[10] approaches. However, while various theoretical studies have demonstrated their high potential to discriminate chiral objects,[11–13] the specific properties of vortex beams interacting with molecules have scarcely been used in spectroscopic applications so far.

The growing experimental interest in HD is triggered by the capability to produce OAM beams in a versatile fashion, for instance using a spatial light modulator. Dichroic effects based on this method have been demonstrated on chiral molecules adsorbed on ordered arrays of plasmonic metallic nanoparticles.[14] However, the interpretation of the signal is complicated by the details of the molecule-nanoparticle interaction. More recently, a giant HD from a single chiral microstructure was reported, reaching 20 % in the visible range.[15] The microstructure had an overall size of over 2 microns with branches having a width of ca. 200-400 nm. However, to date and regardless of the spectral range, there is no experimental demonstration of HD on randomly oriented enantiopure samples, let alone molecules.

In a previous theoretical work, Ye et al.[12] showed how HD is a sensitive probe of molecular chirality, which just like CD, vanishes in randomly oriented achiral samples. In § S1, we review the fundamentals of HD showing how it engages with molecular chirality. The use of the OAM for HD spectroscopy implies some fundamental differences with CD. First, CD is sufficiently well formulated in the magnetic dipole approximation giving rise to the rotational strength, while the electric quadrupole term vanishes upon rotational averaging. On the other hand, HD strongly relies on the spatial relation of the phase and the molecule. As such, it cannot be adequately described in the dipolar approximation. For HD, the multipolar expansion is necessary because several multipoles are needed to accurately describe the electric field near the singularity.[12] To overcome this issue, a non-local formalism was developed based on the minimal coupling Hamiltonian that encompasses all multipoles.[16] In this description, the light-matter coupling is described through an overlap integral of the vector potential and the current density operator (see eq. 1 in the SM). Details on the signal expression are given in § S1. In this description, the helical fields interact with the full transition current densities, a matrix element that indicates the flow of current within the molecule. Second, CD is limited to the difference between left and right circular polarizations, for a SAM of light limited to ±1 values. On the other hand, HD can be carried with all integer ±L values, leading to a variation of signal amplitudes at different wavelengths. This can be understood as follows: transition current densities can be expanded in vector spherical harmonics,[17,18] which also carry a phase twist. The OAM fields with an L-value matching the dominant term into this spherical expansion will yield the strongest signals. In other words, HD is not only a measure of chirality, but it also reveals the degree of chirality by the dependence of the magnitude of the signal difference on the L-value.

As mentioned above, probing the HD of disordered media has serious limitations in the optical domain. On the other hand, the hard X-ray domain offers two major advantages. First, the increased amplitude of the wavevector **k** has the consequence that higher multipoles can contribute to the measured signal. Higher multipoles interact with spatial derivatives of the exciting fields and are thus more sensitive to their spatial variation, i.e. the phase vortex in the present case. Second, the short wavelengths of X-rays allow for much smaller focal spots than achievable in the visible domain, leading to a phase gradient in the illuminated area that is orders of magnitude higher compared to that of visible light. This enhanced phase gradient is favorable, as in actual experiments, the vortex and the chiral compound under investigation are not perfectly aligned: the higher the phase gradient the chiral material experiences, the larger the anticipated interaction cross-section. In contrast, a spatially large vortex locally induces a smaller field gradient, which is expected to lead to a significantly lower HD signal.

In this work, we report the first study of a helicity-dependent spectroscopic signal from an OAM beam in the hard X-ray regime at the Fe K-edge (7.1 keV). We investigate an enantiopure sample of the chiral metal-organic dicationic molecular complex $Fe^{II}$(4,4'-dimethyl-2,2'bipyridine)$_3$ [Fe(4,4'-diMebpy)$_3$]$^{2+}$ shown in Figure 1, with its right and left-handed geometries of Δ or Λ configurations. To control the three-bladed propeller structure, association with chiral anions is used so that the molecule can be stabilized in single Δ or Λ configured salts.[19–21] These chiral anions consist of enantiopure TRISPHAT [tris(tetrachlorobenzendiolato)phosphate(V)] (diastereoselectivity > 96%), whose synthesis and stereo-control are described in § S5.

We conducted the spectroscopic experiments with OAM beams carrying a variety of topological charges. The experiments were carried out at the cSAXS beamline of the Swiss Light Source (Paul Scherrer Institut, Villigen PSI). The experimental set-up is shown in Figure 2. We used diffractive X-ray lenses, spiral Fresnel zone plates (ZP)[22–24] to generate the linearly-polarized OAM-carrying beams. To this aim, we fabricated a set of spiral ZPs with different topological charges in both helicities, including a conventional Fresnel ZP for the case without topological charge. The ZP were prepared with the same lithographic method described in detail in ref. [25]. They are made of gold, and have a diameter of 120 μm, an outermost zone width of 60 nm, and a height of 1.2 μm. The capabilities of these ZPs for generating OAM beams has been demonstrated in previous works.[26–29] To confirm the quality of the vortexed wavefronts, we performed ptychographic scans of an available resolution test sample (Siemens star).[24,28] Figure 3 shows the complex-valued illumination fields at the focal position, which were obtained together with the sample transmissivity in the ptychographic reconstruction and through propagation to the focus. They confirm the quality and geometrical size of the phase vortex created by the spiral ZPs. The L=+3 vortex displays an elongated shape with two local minima which indicates that it is not perfect. Under non-perfect conditions, similar effects have been observed in the visible domain[30]. The molecular samples were placed at the focal position. Fe K-edge X-ray near-edge absorption structure (XANES) spectra were then recorded from 7.10 to 7.15 keV using OAM-carrying beams. Several spectra were acquired for both positive and negative values of the OAM, and for both enantiomers. The XANES spectra were measured in Total Fluorescence Yield (TFY) detection mode using an Eiger 500k detector positioned at 90° from the incoming beam to minimize elastic scattering (see Fig. 2). More details of the experimental procedures are given in § S3.

Figure 4 shows the Fe K-edge XANES spectra of the Λ- and Δ-enantiomers for values of the OAM carried by the beams of ±1 and ±3, as indicated by the L value. The spectra for L=±4 were also recorded but were too noisy due to the limited data acquisition time and are therefore not presented here. The spectra exhibit an overall similar shape with a pre-edge feature at 7.115 keV, an edge feature at 7.125 keV and the main edge peak at 7.131 keV. These features are comparable with already measured and computed XANES spectra[31,32] of racemic $[Fe(bpy)_3]^{2+}$. The pre-edge feature at 7.115 keV is associated with the quadrupolar 1s→3d transition, which becomes partially allowed in the $D_3$ symmetry of the molecular complex. The edge feature at 7.125 keV is attributed to a dipolar 1s→4p transition.[31,33] The 7.131 keV feature represents the onset of the multiple scattering region typical of the above-edge XANES. The XANES spectra have similar amplitudes ($\sim 1 \pm 0.2 \times 10^5$ counts). We note that a slight difference of the XANES amplitude, stemming from variations of the sample thickness, does not alter significantly the asymmetry ratio as the spectra are normalized. Error bars are calculated as type-A uncertainties (see § S6).

It is worth stressing the differences between the +L and -L spectra of the different enantiomers in Figure 4. In particular, for the Δ enantiomer the +L values yield slightly more intense spectra, while the reverse is observed for the Λ enantiomer. This indicates that the transition current density of the Δ enantiomer has an OAM component more important in +L than -L, i.e. its vector multipolar expansion coefficient with momentum +L is larger than the one with momentum -L. The preferred sign for each enantiomer could be related to their three-bladed structure that favours a given sign of the OAM in the multipolar expansion.

The HD signal is defined by the differential absorption of vortex beams carrying ±L OAM and, as for CD spectra, we normalize the signal by the average of the absorption of clockwise and counterclockwise helical beams to define the measured HD asymmetry as:

$$S_{\mathrm{HD}}(L,\omega) = \frac{A_{+L}(\omega) - A_{-L}(\omega)}{\frac{1}{2}[\max A_{+L}(\omega) + \max A_{-L}(\omega)]} \qquad \text{eq. 1}$$

Where $A_{\pm L}(\omega)$ is the XANES spectrum. Unlike the asymmetry ratio commonly used in the optical regime, the maximum of the XANES spectrum is used for normalization. This tends to underestimate the HD within the observed spectral range, but it minimizes the uncertainty of dividing by an almost vanishing signal in the pre-edge region. The HD spectra resulting from Figure 4 measured on enantiopure powders of the Δ- and Λ- enantiomers for L = ±1 and ±3 are displayed in Figure 5, confirming that HD signals exhibit opposite signs for opposite enantiomers, as demonstrated in the SM.

The asymmetry ratio in HD ranges from 1% to 5%, which considering the used normalization, is much higher than what is commonly achieved in standard CD measurements in the optical domain, but also in the very few CD studies in the X-ray domain.[34–38] This is a very promising result in view of applications of HD for the characterization of molecular enantiomers.

The 1s→3d transition in the pre-edge region does not seem to provide a distinct feature, although the HD signal appears to set in this spectral region. Since the HD signal originates from the field spatial variation, it should be sensitive to the quadrupolar couplings within the molecules. Thus, the occurrence of changes in the pre-edge region is possible but needs further investigation. At and above the edge, the sign of the HD signal is opposite for the two enantiomers (positive for Δ and negative for Λ according to our definition) and is the same for the investigated L = ±1 and ±3 OAM. The transition matrix elements entering into the definition of the signal (see § S1) are overlap integrals between field envelope and the transition density currents $\mathbf{j}_{cg}(\mathbf{r})$ linking the ground state g and the core-excited state c. The transition currents provide information on the current needed to move the charges from state g to state c. The phase twist of the field envelope is given by the $e^{iL\varphi}$ term, with φ being the azimuth angle in the plane transverse to the beam propagation. For a given molecule, $\mathbf{j}_{cg}(\mathbf{r})$ is a vector field and the magnitude of each overlap integral, i.e. the strength of the HD beam for a given OAM value, depends on its multipolar expansion. This multipolar expansion is expected to peak for a given L value and HD measurements repeated with many OAMs would provide a multipolar spectrum of the transition current density. In the region above the absorption edge, the spatial shape of the electronic current operator varies slowly since the matter transition matrix elements concerns a transition into the continuum. This explains the sign of the HD spectrum in that region. Additionally, the strength of the HD signal is decreasing towards the end of our acquisition window, i.e. moving into the EXAFS region, in a similar fashion as was observed in X-ray Natural CD (XNCD) experiments.[34–36] For XNCD, it was concluded that single scattering pathways that dominate the EXAFS region cannot lead to chiral sensitivity. This is explained by the fact that single scattering involves only two atoms which always have inversion symmetry, independently of the nearby molecular potential. This reasoning is likely to hold for HD.

The transition matrix elements can also be used to understand the amplitude of the HD signal. It can be expected that the HD signal will be at a maximum for the best geometrical match between the electromagnetic field with a topological charge L and the current density matrix elements $\mathbf{j}_{cg}(\mathbf{r})$. When $\mathbf{j}_{cg}(\mathbf{r})$ is expanded into its multipolar components, the interaction is dominated by the multipoles with the same OAM as the incoming field. The measurement with one sign of L will cancel out the phase in the spherical integral in good approximation and lead to a high transition matrix element while the measurement with the opposite L will add up the phases and result in a smaller overlap integral. In the current experiment, we can observe that the HD signal is about 3 times smaller for measurements with L = ±3 than with L = ±1, indicating that the wavefront gradient of the topological charge L = ±1 fits the molecular geometry better than L = ±3 does. In other words, the asymmetry ratio of helical dichroism as a function of L is a measure of the degree of chirality. More extensive measurements with a larger set of OAM values will permit to obtain an accurate window into the spherical spatial variation of the transition under investigation.

The interest in HD has so far been limited to the optical domain where achieving large field gradients over a molecular size is difficult. This is the reason why only two optical-domain HD studies have successfully been carried out on fairly large plasmonic nanoparticle aggregates[14] or a single microstructure.[15] The specific architecture in ref. [14] generates strong local electric quadrupole fields due to the coupling between adjacent plasmonic nanoparticles. The same reason explains why most experiments in the visible domain use much higher topological charges, typically between 10 and 50. In contrast, we demonstrate herein for the first time HD in a randomly oriented sample consisting of enantiomers of a molecule that can be treated as an isolated entity. Furthermore, this is carried out in the hard X-ray range, which is possible because the higher multipolar contributions are enhanced in the signal. Spiral Fresnel ZPs tuned for X-rays have the added advantage over spiral phase plates commonly used in the optical domain of being achromatic over an extended frequency range, allowing to scan spectra as presented here.

The use of spiral Fresnel ZPs and a ptychography setup opens the way for making HD a standard technique at synchrotron facilities. The enhanced intensity response of HD compared to CD offers very exciting perspectives and a wide range of potential applications. In the present case, the element-sensitivity of X-rays probes the iron atom, at the core (center) of a chiral three-bladed propeller geometry. It is interesting to consider what happens when atoms away from the stereogenic center are interrogated. Simulations[12,39] suggest that the signal would decrease with increasing distance between the probed atom and the stereogenic center. In the case of molecules devoid of stereogenic centers, a probe such as X-ray HD (e.g. at the carbon K-edges of an organic molecule) could provide interesting insights. For example, in propeller structures like the iron complex in this work or screw-like molecules such as helicenes, we can expect a maximum HD with helical beams when the incoming OAM of light matches the handedness and pitch of the molecular screw.

The present work also opens a new window into hard X-ray studies of molecular chirality. Such studies had previously only been achieved by recording CD spectra of crystalline or oriented samples at the K-edge of transition metals.[34–37] The effect was attributed to the interference between allowed electric dipole and electric quadrupole transition moments. Indeed, the electric-magnetic dipole interaction of CD spectroscopy is even excluded for K- or $L_1$-edge spectra, because magnetic dipole transitions are forbidden from s-orbitals, so the only possible source of magnetic dipole intensity involves 1s-2p orbital mixing in addition to core-hole relaxation. The only reported X-ray CD spectrum of a randomly oriented sample concerned gas phase methyloxirane[37] at the carbon K-edge, which was possible because of the 1s-2p orbital mixing, resulting from their small energy separation. The present work demonstrates the feasibility of hard X-ray chirality studies of randomly oriented samples. Additionally, both SAM and OAM can be combined to interact with the molecule, leading to Circular-Helical Dichroism (CHD), a technique that utilizes the differential absorption between pulses with left/right polarization and the wave front helicity. Numerical studies[12] have shown that CHD may lead to even larger observable asymmetry than HD but has the additional challenge of requiring high circular polarization purity and stability. HD measurements can also be implemented in chiral solid materials with magnetic vortices.[40] In this case, the signal will be given as the overlap between the magnetic vortex and the incoming beam phase vortex.[41]

Finally, extension of X-ray HD spectroscopy into the time domain offers exciting perspectives. As time-resolved X-ray absorption spectroscopy has now reached a mature level and is intensely used to investigate chemical and biochemical dynamics,[42,43] the possibility to use X-ray HD to investigate such processes promises major advances in their understanding. In particular, given the enhanced dichroic response in HD, one can envision investigating molecules or quasiparticles that lose, acquire or modify their chirality upon photoexcitation. A recent sub-picosecond deep-ultraviolet CD study of the same complex investigated here was recently reported showing a change of the chirality upon populating of the lowest metastable excited state. This change is caused by a hitherto unknown structural distortion of the organic ligands.[44] Time-resolved HD at the Fe K-edge would provide additional insights and a quantitative estimate of the distortion. Another example is formamide, which is achiral in the ground state and becomes chiral upon photoexcitation due to a double well excited state potential along its bending normal coordinates, with the minima corresponding to the two enantiomers.[45] HD can be used as a probe of this time-evolving chirality and offers a scheme that is not more complicated than CD, yet more insightful given that the OAM can be varied. Such time-resolved HD studies will likely require the time-resolution of X-ray Free Electron Lasers (XFEL) and dedicated engineering to design spiral Fresnel zone plates that are capable of sustaining and taking advantage of the high brilliance of these sources. Adequate material choice to avoid ablation caused by the extremely high flux density at these radiation sources is a possible direction. Spiral wave-fronts have also been generated using a helical undulator at synchrotron[46] or XFEL facilities.[47] Other OAM generation techniques include spiral phase plates,[48] high-harmonic conversion of helical optical beams,[49] which will provide a broad range of alternative to implement steady-state and time-resolved HD.

In summary, in the present work four achievements have been accomplished: the observation of HD in a disordered medium, the first HD study of a molecular system, and the first demonstration of HD in the hard X-ray regime, and for that matter, with element-selectivity. These achievements promise exciting developments on the fundamentals of molecular chirality with element-selectivity, which will undoubtedly pave the way to novel insights into molecular structure and dynamics.

# Acknowledgement

This work was supported by the European Research Council Advanced Grants H2020 ERCEA 695197 DYNAMOX, the Swiss NSF via the NCCR:MUST and grants 200020_169914 and 200021_175649. D.K. acknowledges funding from SNSF under Grant No. 200021_175905. JRR was supported by the Fédération André Marie Ampère (FRAMA) and the LABEX MANUTECH-SISE (ANR-10-LABX-0075) of the Université de Lyon, within the program "Investissements d'Avenir" (ANR-11-IDEX-0007) operated by the French National Research Agency (ANR). CB and GFM were supported via the InterMUST Women Fellowship. GFM acknowledges the support of the European Union's Horizon 2020 research and innovation programme (grant agreement No. 851154). FZ and JL thank the University of Geneva and the Swiss NSF for support via grant 200020-184843.

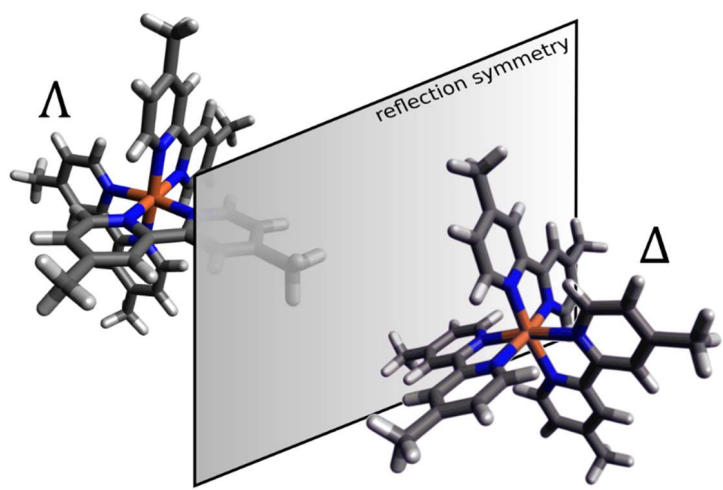

**Figure 1:** Δ (right) and Λ (left) enantiomers of chiral $Fe^{II}$(4,4'-dimethyl-2,2'bipyridine)$_3$ [Fe(4,4'-diMebpy)$_3$]$^{2+}$.

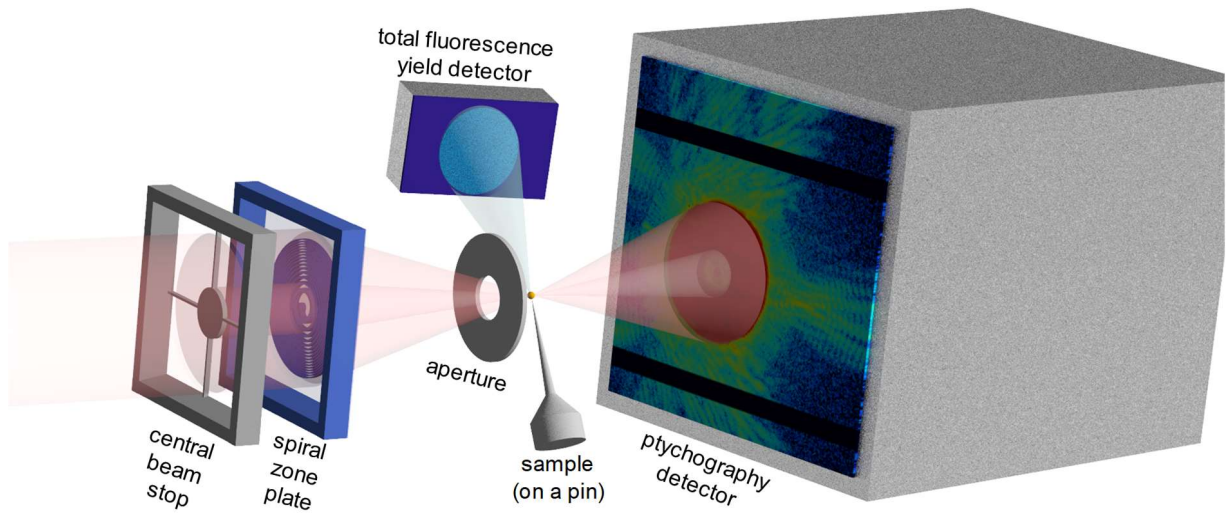

**Figure 2:** Illustration of the experimental setup. The OAM-carrying beam is generated by a spiral zone plate. A central beam stop and an aperture ensure that only the first diffraction order of the zone plate reaches the sample. It is brought into the focus of the beam in solid form on a pin. Spectra are recorded by means of a total fluorescence yield detector that is mounted on the side close to the sample.

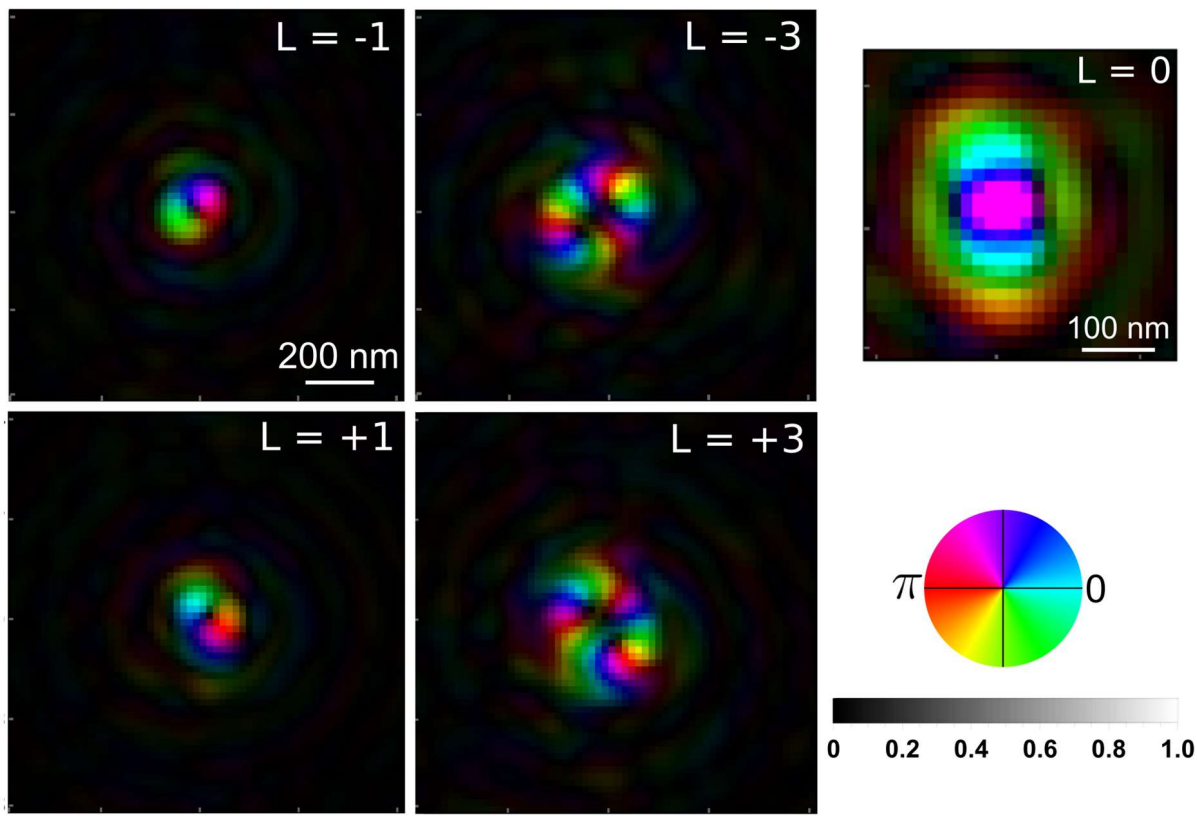

**Figure 3:** Ptychographic reconstructions of helical beams used in the subsequent HD measurements. The OAM of the beam is indicated by the L value in each panel. The colour wheel and the grayscale bar indicate the phase and amplitude in arbitrary units of the field at the focal point.

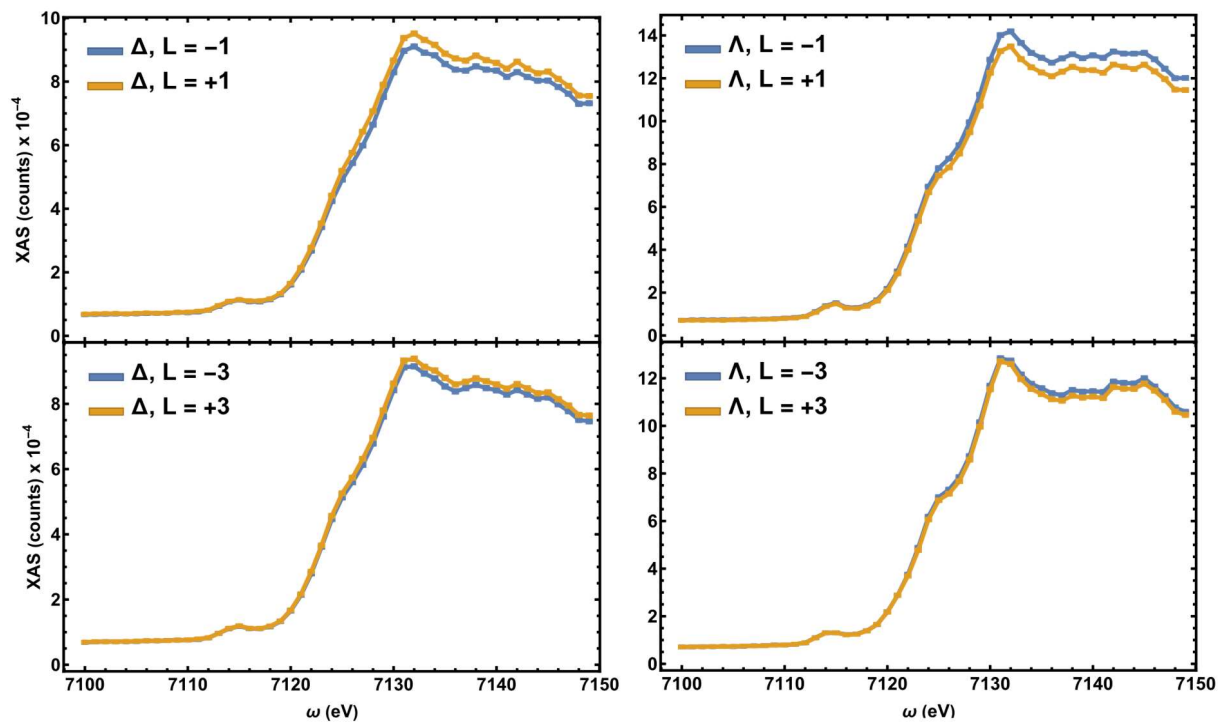

**Figure 4:** Fe K-edge X-ray absorption spectra of $[Fe(4,4'\text{-diMebpy})_3]^{2+}$ with light carrying orbital angular momenta of $L = \pm 1$ and $\pm 3$.

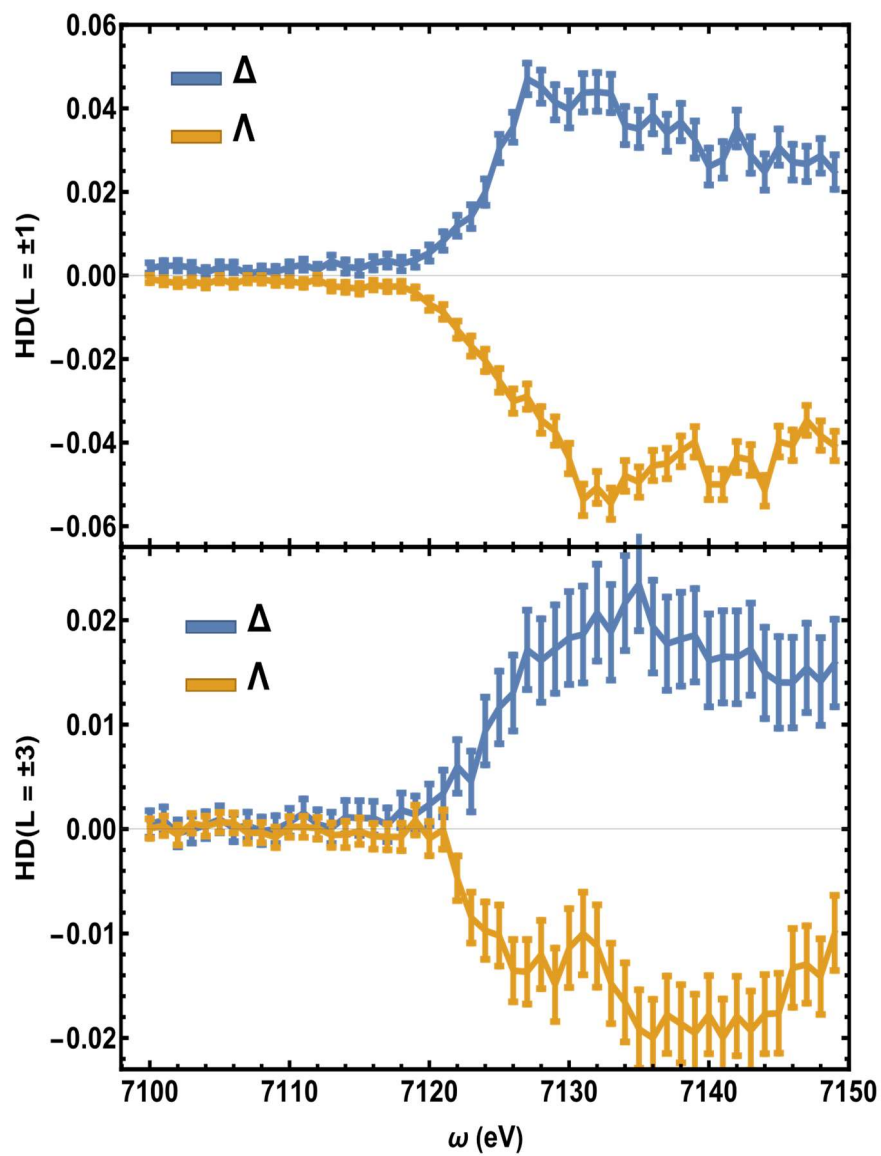

**Figure 5:** Helical dichoism signals at the Fe K-edge of Δ- and Λ-[Fe(4,4'-diMebpy)$_3$]$^{2+}$ for L= ±1 (top) and ±3 (bottom).

# Methods

## Zoneplate fabrication

For fabrication of the spiral zone plates, a 2x2 mm large and 250 nm thick silicon nitride membrane was coated with a conductive metal layer and a 1.4 μm thick polymethyl metacrylate (PMMA) film as lithography resist. After exposure in a 100 kV electron lithography tool (Vistec EBPG 5000+) and development of the exposed parts in a mixture of isopropanole and water (7:3), a 1.2 μm thick gold film was electrochemically deposited. In a subsequent step, the PMMA film was removed with acetone, and the resulting structures where dried in supercritical $CO_2$. With this method, we fabricated spiral zone plates with a diameter of 120 μm, an outermost zone width of 60 nm, and topological charges from L=-9 to L=9 (including conventional Fresnel zone plates without topological charge).

## Ptychography

Ptychography is a diffractive imaging technique that allows to simultaneously reconstruct the amplitude and the phase of both the sample and the illuminating wave.[45,46] This technique is based on proportionality between the diffraction pattern in the far-field and a Fourier transform of the complex-valued transmissivity of the sample. In ptychography the sample is raster scanned with an area-by-area scan in a confined and coherent X-ray beam with a step size smaller than the illumination spot. This results in same areas partially illuminated in multiple scans. Since the probe at each point stays constant, this information can be used to retrieve the wavefront information.[26]

In the current experiment, we used lithographically prepared star patterns (Siemens star) as a sample (see Fig. S5). The general scheme of the setup is shown in Fig. 2. X-ray photons with energy of 7.1 keV are selected by fixed-exit double crystal Si (111) monochromator. The defined illumination on the sample is produced by a central stop (CS), a Fresnel zone plate (FZP), and an order sorting aperture (OSA). To record the scattered X-rays, we used a Pilatus 2M detector positioned 7.252 m from the sample in forward direction. The sample was mounted on a piezo stage allowing accurate positioning with respect to the beam.